\documentclass{ws-procs975x65}

\begin{document}

\title{Four-Body Bound State Formulation in Three-Dimensional Approach \\*(Without Angular Momentum Decomposition)}

\author{M.~R. Hadizadeh$^{\dag}$ and S. Bayegan$^{\ddag}$ }

\address{Department of Physics, University of Tehran, \\
P.O.Box 14395-547, Tehran, Iran\\
E-mail addresses: $^{\dag}$hadizade@khayam.ut.ac.ir ,
$^{\ddag}$bayegan@khayam.ut.ac.ir} \maketitle \abstracts{ The
four-body bound state with two-body forces is formulated by the
Three-Dimensional approach, which greatly simplifies the numerical
calculations of few-body systems without performing the Partial
Wave components. We have obtained the Yakubovsky equations
directly as three dimensional integral equations.}
\section{Introduction}
The four-body bound state calculations are traditionally carried
out by solving coupled Yakubovsky equations in a partial wave
basis. After truncation this leads to two coupled sets of finite
number of coupled equations in three variables for the amplitudes.
This is performed in configuration space[1] and in momentum
space[2,3]. Though a few partial waves often provide qualitative
insight, modern four-body calculations need 1572 or more different
spin, isospin and angular momentum combinations[3]. It appears
therefore natural to avoid a partial wave representation
completely and work directly with vector variables. This is a
common practice in four-body bound state calculations based on
other techniques[4-9]. In recent years W. Gl\"{o}ckle and
collaborators have introduced the three-dimensional approach which
greatly simplifies the two- and three-body scattering and bound
state calculations without using partial wave
decomposition[10-12]. In this paper we extend this approach for
four-body bound state with two-body interactions, we work directly
with momentum vector variables in the Yakubovsky scheme. As a
simplification we neglect spin and isospin degrees of freedom and
treat four-boson bound state. Although the four-boson bound state
has been studied with short-range forces and large scattering
length at leading order in an effective quantum mechanics
approach[13], but it is also based on partial wave approach.
\section{Momentum Space Representation of Yakubovsky Equations in 3-D approach}
The bound state of four identical bosons which interact via
pairwise forces is given by coupled Yakubovsky equations[13]:

\begin{eqnarray}
|\psi_{1}\rangle=G_{0}t_{12}P[(1+P_{34})|\psi_{1}\rangle+|\psi_{2}\rangle]
\nonumber \\*
|\psi_{2}\rangle=G_{0}t_{12}\tilde{P}[(1+P_{34})|\psi_{1}\rangle+|\psi_{2}\rangle]
\label{eq:murn}
\end{eqnarray}
In order to solve coupled equations, $Eq.(1)$, in momentum space
we introduce the four-body basis states corresponding to each
Yakubovsky component:
\begin{eqnarray}
 &&|\vec{u}_{1}\,\vec{u}_{2}\,\vec{u}_{3}\rangle    \nonumber \\
 &&|\vec{v}_{1}\,\vec{v}_{2}\,\vec{v}_{3}\rangle \label{eq:murn}
\end{eqnarray}
Let us now represent coupled equations, $ Eq.(1)$, with respect to
the basis states have been introduced in Eq.(2):
\begin{eqnarray}
 \langle \vec{u}_{1}\,\vec{u}_{2}\,\vec{u}_{3}|\psi_{1}\rangle &=& \int D^{3}u^{'} \int
D^{3}u^{"}\,\langle\vec{u}_{1}\,\vec{u}_{2}\,\vec{u}_{3}|G_{0}t P
|\vec{u}_{1}^{'}\,\vec{u}_{2}^{'}\,\vec{u}_{3}^{'}\rangle \,
\nonumber \\* && \hspace{25mm}
\langle\vec{u}_{1}^{'}\,\vec{u}_{2}^{'}\,\vec{u}_{3}^{'}|
(1+P_{34})|\vec{u}_{1}^{"}\,\vec{u}_{2}^{"}\,
\vec{u}_{3}^{"}\rangle\,\langle\vec{u}_{1}^{"}\,\vec{u}_{2}^{"}\,\vec{u}_{3}^{"}|\psi_{1}\rangle
 \nonumber  \nonumber \\*  &+& \int D^{3}u^{'} \int
D^{3}v^{'}\,\ \langle\vec{u}_{1}\,\vec{u}_{2}\,\vec{u}_{3}|G_{0}t
P |\vec{u}_{1}^{'}\,\vec{u}_{2}^{'}\,\vec{u}_{3}^{'}\rangle \,
\nonumber \\* && \hspace{25mm}
\langle\vec{u}_{1}^{'}\,\vec{u}_{2}^{'}\,\vec{u}_{3}^{'}|\vec{v}_{1}^{'}\,\vec{v}_{2}^{'}\,\vec{v}_{3}^{'}\rangle
\,\langle\vec{v}_{1}^{'}\,\vec{v}_{2}^{'}\,\vec{v}_{3}^{'}|\psi_{2}\rangle
 \nonumber \\*
\langle \vec{v}_{1}\,\vec{v}_{2}\,\vec{v}_{3}|\psi_{2}\rangle &=&
\int D^{3}v^{'} \int D^{3}u^{'}\,
\langle\vec{v}_{1}\,\vec{v}_{2}\,\vec{v}_{3}|G_{0}t \tilde{P}|
\vec{v}_{1}^{'}\,\vec{v}_{2}^{'}\,\vec{v}_{3}^{'}\rangle\,
\nonumber \\* && \hspace{25mm}
\langle\vec{v}_{1}^{'}\,\vec{v}_{2}^{'}\,\vec{v}_{3}^{'}|
(1+P_{34})|\vec{u}_{1}^{'}\,\vec{u}_{2}^{'}\,\vec{u}_{3}^{'}\rangle\langle\vec{u}_{1}^{'}\,\vec{u}_{2}^{'}\,\vec{u}_{3}^{'}|\psi_{1}\rangle
  \nonumber \\*  &+& \int
D^{3}v^{'}\,\langle\vec{v}_{1}\,\vec{v}_{2}\,\vec{v}_{3}|G_{0}t
\tilde{P}|
\vec{v}_{1}^{'}\,\vec{v}_{2}^{'}\,\vec{v}_{3}^{'}\rangle
\vec{v}_{1}^{'}\,\vec{v}_{2}^{'}\,\vec{v}_{3}^{'}\rangle\langle\vec{v}_{1}^{'}\,\vec{v}_{2}^{'}\,\vec{v}_{3}^{'}|\psi_{2}\rangle
 \label{eq:murn}
\end{eqnarray}
Where $D^{3}A=d^{3}A_{1}\,d^{3}A_{2}\,d^{3}A_{3}$. After
evaluating the following matrix elements:
\begin{eqnarray}
\langle\vec{u}_{1}\,\vec{u}_{2}\,\vec{u}_{3}|G_{0}t P
|\vec{u}_{1}^{'}\,\vec{u}_{2}^{'}\,\vec{u}_{3}^{'}\rangle &=&
\frac{\delta^{3}(\vec{u}_{3}-\vec{u}_{3}^{'})}{E-\frac{u_{1}^{2}}{m}
-\frac{3u_{2}^{2}}{4m}-\frac{2u_{3}^{2}}{3m}} \nonumber
\\* &&   \{ \,\, \delta^{3}(\vec{u}_{2}
+\vec{u}_{1}^{'}+ \frac{1}{2}\vec{u}_{2}^{'} )\,
\langle\vec{u}_{1}|t(\epsilon) |\frac{1}{2}\vec{u}_{2}
+\vec{u}_{2}^{'}\rangle \nonumber
\\* && \,\,+ \delta^{3}(\vec{u}_{2}
-\vec{u}_{1}^{'}+ \frac{1}{2}\vec{u}_{2}^{'} )\,
\langle\vec{u}_{1}|t(\epsilon) |\frac{-1}{2}\vec{u}_{2}
-\vec{u}_{2}^{'}\rangle  \} \nonumber
\\*
\langle\vec{v}_{1}\,\vec{v}_{2}\,\vec{v}_{3}|G_{0}t \tilde{P}
|\vec{v}_{1}^{'}\,\vec{v}_{2}^{'}\,\vec{v}_{3}^{'}\rangle &=&
\frac{\delta^{3}(\vec{v}_{2}+\vec{v}_{2}^{'})\,
\delta^{3}(\vec{v}_{3}-\vec{v}_{1}^{'}) }{E-\frac{v_{1}^{2}}{m}
-\frac{v_{2}^{2}}{2m}-\frac{v_{3}^{2}}{m}} \,\,
\langle\vec{v}_{1}|t(\epsilon^{*}) |\vec{v}_{3}^{'}\rangle
\nonumber \\*
\langle\vec{u}_{1}^{'}\,\vec{u}_{2}^{'}\,\vec{u}_{3}^{'}|
(1+P_{34})|\vec{u}_{1}^{"}\,\vec{u}_{2}^{"}\,
\vec{u}_{3}^{"}\rangle &=&
\delta^{3}(\vec{u}_{1}^{'}-\vec{u}_{1}^{"})\nonumber \\* &\times&
\{\,\,
\delta^{3}(\vec{u}_{2}^{'}-\vec{u}_{2}^{"})\,\delta^{3}(\vec{u}_{3}^{'}-\vec{u}_{3}^{"})
\nonumber \\* && \,\, \,+  \delta^{3}(\vec{u}_{2}^{'}
-\frac{1}{3}\vec{u}_{2}^{"} -\frac{8}{9}\vec{u}_{3}^{"} )\,
\delta^{3}(\vec{u}_{3}^{'} -\vec{u}_{2}^{"}
+\frac{1}{3}\vec{u}_{3}^{"} )\,\} \nonumber \\*
\langle\vec{v}_{1}^{'}\,\vec{v}_{2}^{'}\,\vec{v}_{3}^{'}|
(1+P_{34})|\vec{u}_{1}^{'}\,\vec{u}_{2}^{'}\,
\vec{u}_{3}^{'}\rangle &=&
\delta^{3}(\vec{u}_{1}^{'}-\vec{v}_{1}^{'})\nonumber \\* &\times&
\{\,\, \delta^{3}(\vec{u}_{2}^{'}+\frac{2}{3}\vec{v}_{2}^{'}-
\frac{2}{3}\vec{v}_{3}^{'} )\,\
\delta^{3}(\vec{u}_{3}^{'}+\frac{1}{2}\vec{v}_{2}^{'}+\vec{v}_{3}^{'})
\nonumber
\\* && \,\, \,+  \delta^{3}(\vec{u}_{2}^{'}+\frac{2}{3}\vec{v}_{2}^{'}+
\frac{2}{3}\vec{v}_{3}^{'} )\,\
\delta^{3}(\vec{u}_{3}^{'}+\frac{1}{2}\vec{v}_{2}^{'}-\vec{v}_{3}^{'})\,\}
\end{eqnarray}

We can rewrite the coupled equations, $Eq.(3)$, as below coupled
three-dimensional integral equations:
\begin{eqnarray}
\langle \vec{u}_{1}\,\vec{u}_{2}\,\vec{u}_{3}|\psi_{1}\rangle &=&
\frac{1}{{E-\frac{u_{1}^{2}}{m}
-\frac{3u_{2}^{2}}{4m}-\frac{2u_{3}^{2}}{3m}}}\int d^{3}u_{2}^{'}
\,\, \langle\vec{u}_{1}|t_{s}(\epsilon) |\frac{1}{2}\vec{u}_{2}
+\vec{u}_{2}^{'}\rangle \, \nonumber \\* &\times& \{\,\,
\langle\vec{u}_{2}+\frac{1}{2} \vec{u}_{2}^{'} \,\,
\vec{u}_{2}^{'}\,\,\vec{u}_{3}|\psi_{1}\rangle \nonumber \nonumber
\\* \quad && \hspace{2mm} +\langle\vec{u}_{2}+\frac{1}{2}
\vec{u}_{2}^{'} \,\, \frac{1}{3}\vec{u}_{2}^{'}+
\frac{8}{9}\vec{u}_{3} \,\,
\vec{u}_{2}^{'}-\frac{1}{3}\vec{u}_{3}|\psi_{1}\rangle
 \nonumber \nonumber \\*  && \hspace{2mm} +
 \langle\vec{u}_{2}+\frac{1}{2} \vec{u}_{2}^{'}\,\, -\vec{u}_{2}^{'}-\frac{2}{3}\vec{u}_{3} \,\,
\frac{1}{2}\vec{u}_{2}^{'}-\frac{2}{3}\vec{u}_{3}
|\psi_{2}\rangle\,\, \}
 \nonumber \\*
\langle \vec{v}_{1}\,\vec{v}_{2}\,\vec{v}_{3}|\psi_{2}\rangle &=&
\frac{\frac{1}{2}}{E-\frac{v_{1}^{2}}{m}
-\frac{v_{2}^{2}}{2m}-\frac{v_{3}^{2}}{m}} \,\,  \int
d^{3}v_{3}^{'} \, \langle\vec{v}_{1}|t_{s}(\epsilon^{*})|
\vec{v}_{3}^{'}\rangle\, \nonumber \\* &\times& \{\,\, 2\,
\langle\vec{v}_{3}\,\,
 \frac{2}{3}\vec{v}_{2}+\frac{2}{3}\vec{v}_{3}^{'} \,\, \frac{1}{2}\vec{v}_{2}-\vec{v}_{3}^{'} |\psi_{1}\rangle
  \nonumber \\*  && \hspace{2mm}+
\langle\vec{v}_{3}\,\,-\vec{v}_{2}\,\,
\vec{v}_{3}^{'}|\psi_{2}\rangle \,\,\}
 \end{eqnarray}

Here $\epsilon$ and $\epsilon^{*}$ are two-body subsystem energies
and $\langle\vec{a}|t_{s}(\varepsilon)| \vec{b}\rangle\,$ is the
symmetrized two-body $t$-matrix[10]. The so obtained Y-amplitudes
fulfill the below symmetry relations, as can be seen from
$Eq.(5)$.
\begin{eqnarray}
\langle \vec{u}_{1}\,\vec{u}_{2}\,\vec{u}_{3}|\psi_{1}\rangle &=&
\langle -\vec{u}_{1}\,\vec{u}_{2}\,\vec{u}_{3}|\psi_{1}\rangle
\nonumber \\*
\langle\vec{v}_{1}\,\vec{v}_{2}\,\vec{v}_{3}|\psi_{2}\rangle
\nonumber &=&
\langle-\vec{v}_{1}\,\vec{v}_{2}\,\vec{v}_{3}|\psi_{2}\rangle
\nonumber
\\* \langle\vec{v}_{1}\,\vec{v}_{2}\,\vec{v}_{3}|\psi_{2}\rangle
&=& \langle\vec{v}_{1}\,\vec{v}_{2}\,-\vec{v}_{3}|\psi_{2}\rangle
\end{eqnarray}

\section{Choosing Coordinate Systems}
The Y-components $|\psi_{i}(\vec{a}\,\,\vec{b}\,\,\vec{c} )
\rangle$ are given as function of Jacobi momenta vectors as
solution of coupled three-dimensional integral equations,
$Eq.(5)$. Since we ignore spin and isospin dependencies, for the
ground state both Y-components
$|\psi_{i}(\vec{a}\,\,\vec{b}\,\,\vec{c})\rangle$ are scalars and
thus only depend on the magnitudes of Jacobi momenta and the
angles between them. The first important step for an explicit
calculation is the selection of independent variables, one needs
six independent variables to uniquely specify the geometry of the
three vectors[12]. Therefore in order to solve $Eq.(5)$ directly
without introducing partial wave projection, we have to define
suitable coordinate systems. For both Y-components we choose the
third vector parallel to $Z-$axis, the second vector in the $X-Z$
plane and express the remaining vectors, the first as well as the
integration vectors, with respect to them. With this choice of
variables we can obtain the explicit representation for the
Y-components $|\psi_{1}\rangle $ and $|\psi_{2}\rangle $ as [14]:

\begin{eqnarray}
\psi_{1}(u_{1}\,u_{2}\,u_{3}\,x_{1}\,x_{2}\, x_{12}^{3} ) &=&
\frac{1}{{E-\frac{u_{1}^{2}}{m}
-\frac{3u_{2}^{2}}{4m}-\frac{2u_{3}^{2}}{3m}}}\int d^{3}u_{2}^{'}
\,\, t_{s}(u_{1},\tilde{\pi},\tilde{x} ;\epsilon) \, \nonumber \\*
&\times& \{\,\, \psi_{1}(\pi_{1}\,\, u_{2}^{'}\,\, {u}_{3}\,\,
x_{12}\,\,x_{13}\,\,x_{\pi_{1}u_{2}'}^{u_{3}}) \nonumber
\\* \quad && \hspace{2mm} +\psi_{1}(\pi_{1}\,\, \pi_{2}\,\, \pi_{3}\,\,
x_{22}\,\,x_{23}\,\,x_{\pi_{1}\pi_{2}}^{\pi_{3}})
 \nonumber \\*  && \hspace{2mm} +
 \psi_{2}(\pi_{1}\,\, \pi_{4}\,\, \pi_{5}\,\,
x_{32}\,\,x_{33}\,\,x_{\pi_{1}\pi_{4}}^{\pi_{5}})\,\,\}
 \nonumber \\*
\psi_{2}(v_{1}\,v_{2}\,v_{3}\,X_{1}\,X_{2}\, X_{12}^{3} ) &=&
\frac{\frac{1}{2}}{E-\frac{v_{1}^{2}}{m}
-\frac{v_{2}^{2}}{2m}-\frac{v_{3}^{2}}{m}} \,\,  \int
d^{3}v^{'}_{3} \, t_{s}(v_{1},v_{3}^{'},Y_{13^{'}};\epsilon^{*})
 \nonumber \\* &\times& \{\,\, 2\,
\psi_{1}(v_{3}\,\,\Sigma_{1} \,\,\Sigma_{2} \,\,
X_{12}\,\,X_{13}\,\,X_{v_{3}\Sigma_{1}}^{\Sigma_{2}})
  \nonumber \\*  && \psi_{2}(v_{3}\,\,v_{2}
\,\,v_{3}' \,\, X_{22}\,\,X_{23}\,\,X_{v_{3}v_{2}}^{v_{3}'})
\,\,\}
 \end{eqnarray}
The above coupled three-dimensional integral equations are the
starting point for numerical calculations.

\section{Summary}
An alternative approach for four-body bound state calculations,
which are based on solving the coupled Y-equations in a partial
wave basis, is to work directly with momentum vector variables. We
formulate the coupled Y-equations for identical particles as
function of vector Jacobi momenta, specifically the magnitudes of
the momenta and the angles between them. We expect that coupled
three-dimensional Y-equations can be handled in a straightforward
and numerically reliable fashion.
\section*{Acknowledgments}
One of authors (M. R. H.)would like to thank H. Kamada and Ch.
Elster for helpful discussions during EFB19 and APFB05
conferences.

\end{document}